\documentclass[aip,amsmath,amsfonts,amssymb,reprint,nofootinbib]{revtex4-1}

\usepackage{graphicx}  
\usepackage{dcolumn}   
\usepackage{bm}        
\usepackage{amssymb}   
\usepackage{amsmath}

\usepackage{float}
\usepackage{mwe} 
\usepackage{hyperref}
\hypersetup{colorlinks,allcolors=blue}

\usepackage{epsfig}
\usepackage{textcomp}
\usepackage{blkarray}

\usepackage[export]{adjustbox}
\usepackage{textcomp}
\usepackage{color}
\usepackage[english]{babel} 
\usepackage{blindtext}
\usepackage{soul}
\usepackage{xcolor}
\usepackage{bm} 

\begin{document}



\title{ An information-theoretic approach to infer the underlying interaction
	domain among elements from finite length trajectories in a noisy
	environment}

\author{Udoy~S.~Basak$\dagger$}{\let\thefootnote\relax\footnote{{$\dagger$: Equally contributed.}}}
\affiliation{Graduate School of Life Science, Transdisciplinary Life Science Course, Hokkaido University, Kita 12, Nishi 6, Kita-ku, Sapporo 060-0812, Japan}
\affiliation{Pabna University of Science and Technology, Pabna 6600, Bangladesh}%

\author{Sulimon Sattari$\dagger$}
\affiliation{Research Center of Mathematics for Social Creativity, Research Institute for Electronic Science, Hokkaido University, Kita 20, Nishi 10, Kita-ku, Sapporo 001-0020, Japan}

\author{Hossain M. Motaleb}
\affiliation{Research Center of Mathematics for Social Creativity, Research Institute for Electronic Science, Hokkaido University, Kita 20, Nishi 10, Kita-ku, Sapporo 001-0020, Japan}
\affiliation{University of Dhaka, Dhaka 1000, Bangladesh}%

\author{Kazuki Horikawa}
\affiliation{Department of Optical Imaging, The Institute of Biomedical Sciences, Tokushima University Graduate School, 3-18-15 Kuramoto-cho, Tokushima City, Tokushima, 770-8503, Japan}

\author{Tamiki Komatsuzaki\thanks{tamiki@es.hokudai.ac.jp}}
\affiliation{Graduate School of Life Science, Transdisciplinary Life Science Course, Hokkaido University, Kita 12, Nishi 6, Kita-ku, Sapporo 060-0812, Japan}
\affiliation{Research Center of Mathematics for Social Creativity, Research Institute for Electronic Science, Hokkaido University, Kita 20, Nishi 10, Kita-ku, Sapporo 001-0020, Japan}
\affiliation{Institute for Chemical Reaction Design and Discovery (WPI-ICReDD), Hokkaido University Kita 21 Nishi 10, Kita-ku, Sapporo, Hokkaido 001-0021, Japan}

\affiliation{Graduate School of Chemical Sciences and Engineering Materials Chemistry and Engineering Course, Hokkaido University, Kita 13, Nishi 8, Kita-ku, Sapporo 060-0812, Japan}
%
\vskip 0.25cm

\begin{abstract}
Transfer entropy in information theory was recently demonstrated [Phys. Rev. E 102, 012404 (2020)] to enable us to elucidate the interaction domain among interacting elements solely from an ensemble of trajectories. There, only pairs of elements whose distances are shorter than some distance variable, termed cutoff distance, are taken into account in the computation of transfer entropies. The prediction performance in capturing the underlying interaction domain is subject to noise level exerted on the elements and the sufficiency of statistics of the interaction events. In this paper, the dependence of the prediction performance is scrutinized systematically on noise level and the length of trajectories by using a modified Vicsek model. The larger the noise level and the shorter the time length of trajectories, the more the derivative of average transfer entropy fluctuates, which makes it difficult to identify the interaction domain in terms of the position of global minimum of the derivative of average transfer entropy. A measure to quantify the degree of strong convexity at coarse-grained level is proposed. It is shown that the convexity score scheme can identify the interaction distance fairly well even while the position of global minimum of the derivative of average transfer entropy does not.  We also derive an analytical model to explain the relationship between the interaction domain and the change of transfer entropy that supports our cutoff distance technique to elucidate the underlying interaction domain from trajectories.
  \end{abstract}

\maketitle
\section{Introduction}
Collective migration is the synchronized movement of agents emerging from the mutual interactions between them \cite{grada2017research,theveneau2017leaders}.  One of the basic properties of collective motion is that the movement of an individual is influenced by the movement of other individuals in its local vicinity and/or through long range interactions, e.g., via some signals such as chemicals emitted by cells.   At the cellular level, collective motion can be observed in wound healing, cancer development, and organogenesis \cite{friedl2009collective, haeger2015collective,trepat2009physical}. The question of how microscopic 
interactions between agents regulate the macroscopic group behavior is one of the most intriguing subjects \cite{lord2016inference}. This is closely related to the problem of causal inference within systems composed of many agents.\par

For the qualitative understanding of collective motion, a variety of simulation models have been proposed such as the Reynolds'  flocking model \cite{reynolds1987flocks}, the Vicsek model (VM) \cite{PhysRevLett.75.1226}, and the Couzin model \cite{couzin2002collective}. Among these,  
the VM has been widely used to study various dynamics of collectively moving self-propelled particles, such as symmetry breaking \cite{PhysRevE.77.021920,creppy2016symmetry}, phase transition \cite{PhysRevLett.75.1226,PhysRevE.74.061908}, and classification of leaders and followers  \cite{PhysRevE.93.042411,mwaffo2017analysis}. In the VM, each particle moves with a constant speed and  its direction of motion is determined by the average direction of motion of its neighboring particles in the presence of noise \cite{PhysRevLett.98.095702,liu2008connectivity,chate2008modeling}. In other words, a moving particle interacts only with the particles within a distance of $R$ as it would via direct interactions but also via signal transduction such as chemicals. It is noted that the interaction between pairs of particles is not reciprocal to mimic the self-organized, collective behavior.   \par

One of the possible drivers of collectively moving agents is the presence of influential individuals, sometimes referred to as `leaders', who control the movement of the other individuals, referred to as the `followers'. Leader-follower relationships have been studied in a fish shoal  \cite{krause2000leadership},  troops of baboons  \cite{sueur2011group}, in a colony of honey bee  \cite{seeley2009wisdom} and so forth. At the cellular level, it has been studied that collective migration of MDCK epithelial cells \cite{yamaguchi2015leader,reffay2011orientation}, wound healing \cite{omelchenko2003rho}, cancer growth in  breast \cite{cheung2013collective}, etc.,  are regulated by the leader cells.\par
     
Identifying leader and follower agents is a challenging endeavor. First and foremost, one must identify what it means to be a leader. Based on asymmetric nature of influence on activities among entities, in this article, we define a `leader' as an entity which more influences (on average) on the activity of the other entities (termed `followers'). Once leadership has been defined, various types of empirical data, e.g.,  ensembles of trajectories of agents, can be used to infer the differential influence in interaction and identify leader-follower relationships. By definition, leaders are expected to be more persuasive compared to the followers.  Since followers follow the movement of leaders, some correlation should exist between some physical quantity related to a leader and that related to a follower with a certain time delay, as information cannot travel from a leader to a follower at infinite speed. \par

To measure causal influence among multivariate time series, and also to classify different types of particles in a multiple-particle system, information theory provides a variety of approaches \cite{hlavavckova2007causality,green2013relationship}. Some of the typical quantities used are mutual information \cite{cover2012elements}, time-delayed mutual information \cite{PhysRevLett.85.461}, transfer entropy \cite{PhysRevLett.85.461}, and causation entropy \cite{sun2014causation}. These quantities have  been computed using time lapse motion data of moving individuals to determine the directions of influence. For example, it was found by using a zebrafish interaction model that  net transfer entropy is a more accurate classifier than extreme-event synchronization and cross-correlation for classifying leaders and followers \cite{PhysRevE.93.042411}. In swarms of bats,  transfer entropy was used to demonstrate that there exists a leader-follower relationship between the front bat and the rear bat \cite{orange2015transfer}. Using the trajectories of handball players, it was showed that transfer entropy is capable of capturing  the causal relationships between players \cite{itoda2015model}. 
 \par 

In the above-mentioned studies, all pairs of agents are taken into account at every time instance to evaluate transfer entropy irrespective of the distance between the agents. 
This is not necessarily an optimal use of the data available for capturing the underlying leader-follower relationship, given that the interaction domain is known. 
 It was shown, using a modified VM, that the classification scores of leaders and followers significantly increase upon incorporating the identified interaction domain information compared to the conventional way of transfer entropy estimation where the distance between the agents is not taken into account \cite{PhysRevE.102.012404}. When two particles come into their interaction domain, they may share or transfer information which results in some change in their motion such as the direction of motion. As the distance between them exceeds the interaction radius, the amount of information flow decreases and goes to zero at the limit of the distance being infinity in a fluctuating environment. This methodology requires that the interaction domain is known, which may not be the case. A new scheme has been proposed to estimate the underlying interaction domain from the trajectories of particles to monitor the change in transfer entropy as a function of the distance between them, called cutoff distance $\lambda$.  It was demonstrated that the derivative of average transfer entropy (and also cross correlation) with respect to $\lambda$ has a minimum near the interaction domain, by which one can identify the underlying interaction domain from a set of trajectories \cite{PhysRevE.102.012404}. \par 

The scheme is dependent on how transfer entropy can be estimated so that it takes into account enough statistics of interacting particles, and
positions and numbers of the minimum of the derivative of  average transfer entropy along the cutoff distance $\lambda$ may also be subject to the extent of external noise and time length of trajectories. In this paper, we scrutinize how the prediction performance in capturing the underlying interaction domain depends on the size of noise and time length of the trajectory data. We also examine an alternative scheme expected to be stable against noises and time length, that relies on the degree of convexity at coarse-grained scale in the derivative of average transfer entropy along the cutoff distance, and time variance of underlying interaction radius of particles.          
  
 \par

 \section{Identification of Leaders and followers}
  
Transfer Entropy (TE) from time series of a stochastic variable $X=\{...,x_{t-1},x_t,x_{t+1},...\}$ to time series of another stochastic variable $Y=\{...,y_{t-1},y_t,y_{t+1},...\}$ is defined
as \cite{PhysRevLett.85.461}:
\begin{align}
\text{TE}_{X\to Y}
&=I(y_{t+\tau};x_t|y_t),\nonumber \\
&=\sum_{y_{t+\tau},y_t,x_t}p(y_{t+\tau},y_{t},x_t)\log_2 (\frac{p(y_{t+\tau}|y_t,x_t)}{p(y_{t+\tau}|y_t)}),\nonumber \\
&= H(y_{t+\tau}|y_t)-H(y_{t+\tau}|y_t,x_t) ,\label{eqn. TE}
\end{align}
where $\tau$ is the time lag between the two time instants and $H(.|.)$ represents the conditional Shannon entropy \cite{cover2012elements}.
 TE is proven to be non-negative.  
 A positive value of $\text{TE}_{X \to Y}$ is considered to indicate the causal influence of $X$ on $Y$ \cite{PhysRevLett.116.238701}. For a pair of agents $X$ and $Y$, the net transfer entropy from $X$ to $Y$, defined as  $\text {NTE}_{X \to Y}=\text{TE}_{X \to Y}-\text{TE}_{Y\to X}$ can be used to infer the direction of causal influence. A positive $\text{NTE}_{X\to Y}$ may indicate that $Y$ follows $X$, which quantifies the causal direction from $X$ to $Y$.
 
 As a classifier to differentiate leaders and followers, the average net transfer entropy is used, which is denoted as $\chi$ and defined for a given particle $i$ as follows:
 \begin{equation*}
 \chi^{(i)}=\frac{1}{N-1}\sum_{j(\neq i)}(\text{TE}_{i \to j}-\text{TE}_{j\to i}),
 \end{equation*}
 where $\text{TE}_{i \to j}$ represents TE from the particle $i$ to $j$ and $N$ is the total number of particles in the system. The value of $\chi^{(i)}$ for each particle $i$ is compared to a selected threshold value $\epsilon$. A particle for which $\chi^{(i)}$ is higher than the threshold $\epsilon$ is identified as a leader, otherwise it is identified as a follower. The resulting classification is compared to the ground truth to determine how many leaders (followers) are identified correctly. Based on these statistics, the true-positive rate and the false-positive rate are computed for the chosen $\epsilon$ as follows \cite{hajian2013receiver}\newline
 \begin{align}
 \text{True-Positive Rate}&=\frac{\text{True positive}}{\text{True positive+False negative}},\nonumber \\
 \text{False-Positive Rate}&=\frac{\text{False positive}}{\text{False positive+True negative}}.\nonumber
 \end{align}
 
To show the classification performance of a classifier receiver-operating characteristic 
curve is used.   It is obtained by plotting the true positive rate versus false positive rate at different  values of $\epsilon$ \cite{PhysRevE.93.042411}.   
 In order to quantify the accuracy of the classifier\textquotesingle s performance and to compare the performance of different classifiers, area under receiver-operating characteristic curve (AUC) has been used \cite{hanley1982meaning}. An AUC score of 1.0 means that that classifier accurately predicts the identities of the particles whereas a value of 0.5 means that the classifier has no class separation capacity whatsoever.  \par

\section{Modified Vicsek Model}
Similar to the standard VM \cite{PhysRevLett.75.1226}, we consider that $N$ self-propelled particles are moving with the same constant speed $v_0$ in  a two-dimensional square box of length $L$ with  periodic boundary conditions, and at time $t=0$ the particles are positioned and oriented randomly. At time $t+1$,  the position of $i$th particle is denoted by $\vec{r}_i^{t+1}$  is updated  at each time step $\Delta t$ as:
\begin{equation}
\label{eqn: position}
\vec{r}_i^{t+1}=\vec{r}_i^t+\vec{v}_i^t \Delta t,
\end{equation} 

\noindent where $\vec{r}_i^t$ denotes the  position of  $i$th particle at time $t$, and $\vec{v}_i^{t}$ represents the corresponding velocity of the particle with an absolute speed $v_0$ and a direction given by the angle $\theta_i(t)$. This angle is obtained from the following equation:
\begin{align}
\label{eqn: weight}
\theta_i(t+1)
&= \langle {\bm \theta}(t)\rangle_{R,\textbf{\textit{w}},\vec{r}_i^t}+\Delta\theta_i.
\end{align}
\indent Here $\langle\bm{\theta}(t)\rangle_{R,\textbf{\textit{w}},\vec{r}_i^t} $ is the weighted orientation averaged over particles (including the particle $i$ itself), which are within a circle of radius $R$ centered on the position  $\vec{r}_i^t$ of the particle $i$ at time $t$, computed by  $\arctan\left[\sum'_{j}w_{ji}\sin\theta_j(t)/\sum'_{j}w_{ji}\cos\theta_j(t)\right]$ where $\sum'$ takes over all $j$ satisfying $\mid\vec{r}_i^t-\vec{r}_j^t\mid\le R$ \cite{PhysRevE.102.012404}. $\textbf{\textit{w}}$ is a matrix whose element $w_{ij}$ corresponds to the interaction strength that the particle $i$ exhibits on its neighboring particle $j$.  If the particle $i$ is a leader and $j$ is a follower, then $w_{ij}>w_{ji}$. Also the interaction strength of a particle $i$ on itself is 1.0 i.e., $w_{ii}=1.0$. We set the values of leaders\textquotesingle~ and followers\textquotesingle~   interaction strengths to 1.05 and 1.00, respectively. The greater the difference between $w_{\text{LF}}$ and $w_{\text{FL}}$ the easier leader and follower particles are classified. Thus,  higher $w_{\text{LF}}$ would produce higher AUC scores irrespective of the cutoff distance we would employ. Hence to study the effect of cutoff distance on the classification score, a just slightly higher value is chosen for $w_{\text{LF}}$ compared to $w_{\text{FL}}$. $\Delta \theta_i$ represents random number at time $t$ which can be chosen with a uniform probability distribution from the interval $[ -\eta_0/2,\eta_0/2]$, where $\eta_0$ may be considered as a temperature-like parameter.
The total time length is designated by $T$ during which transfer entropies are estimated between leader and follower particles. \par

\begin{figure}[hbt]
	\centering
	\includegraphics[width=.9\linewidth]{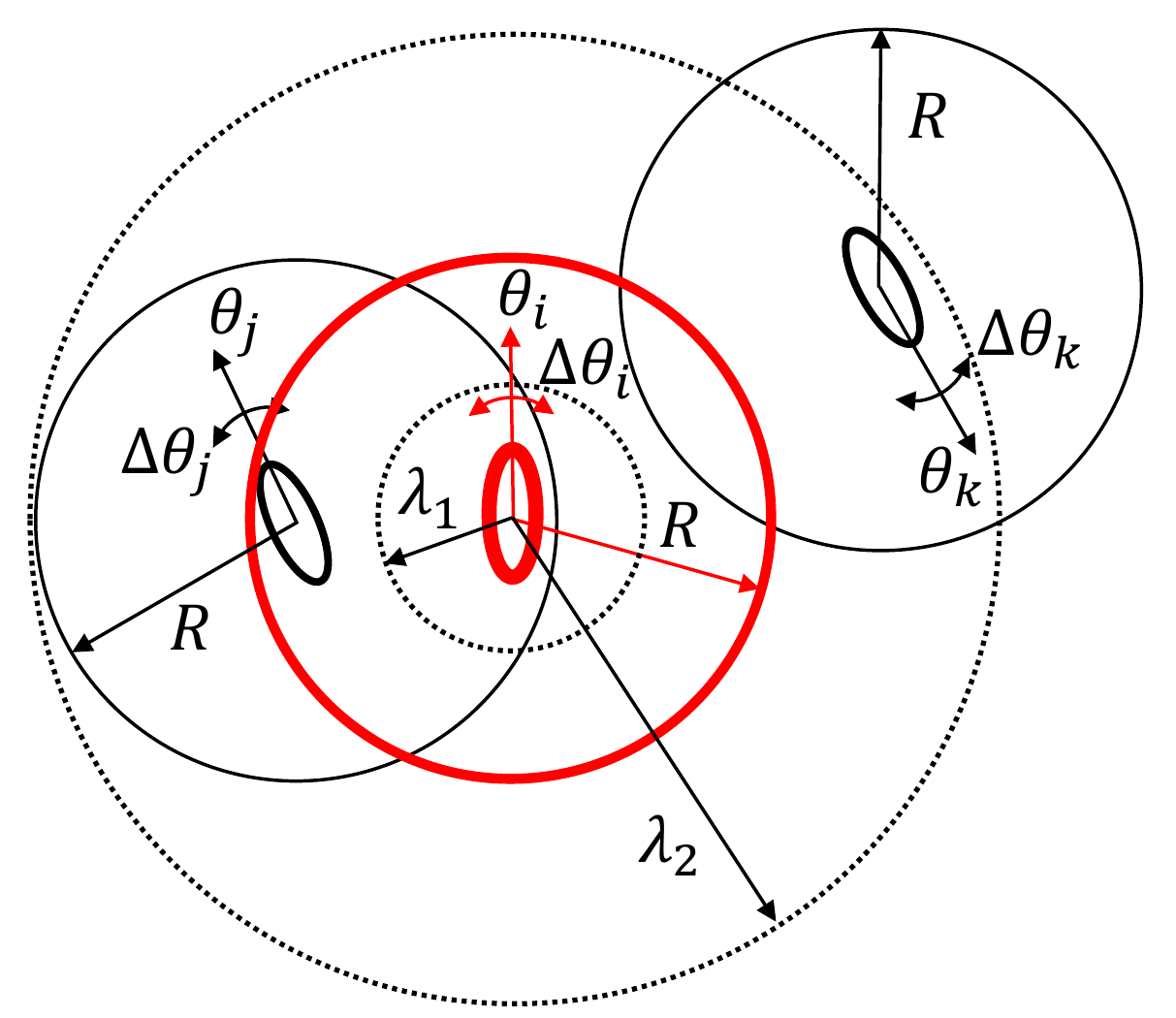}
	\caption{Schematic diagram of cutoff distance. Ovals represent moving particles. $\theta_i$, $\theta_j$, and $\theta_k$ represent the direction of motion of receptive particles at time $t$ and $R$ is the interaction radius. $\lambda_1$ and $\lambda_2$ exemplify two different cutoff distances.  For example, for the cutoff distance $\lambda=\lambda_1$, the actual distance between the particles $i$ and $j$ at this time instance is larger than $\lambda_1$, and hence, their orientation information is not considered for TE calculation between   them even though the particles are located within each other's interaction domain (likewise the orientational information between the particle $i$ and $k$ is not taken into account in the computation of TE at $\lambda=\lambda_1$). But for $\lambda=\lambda_2$, particles $j$ and $k$ are both located within the distance of $\lambda_2$ from particle $i$.    Hence the orientation information of $\theta_k$ and $\theta_i$ and that of $\theta_j$ and $\theta_i$ are considered to compute TE irrespective of the underlying interaction radius $R$.}
	\label{fig: cutoff_cartoon}
\end{figure}
\begin{figure}[htb]
	\centering
	\includegraphics[width=.9\linewidth]{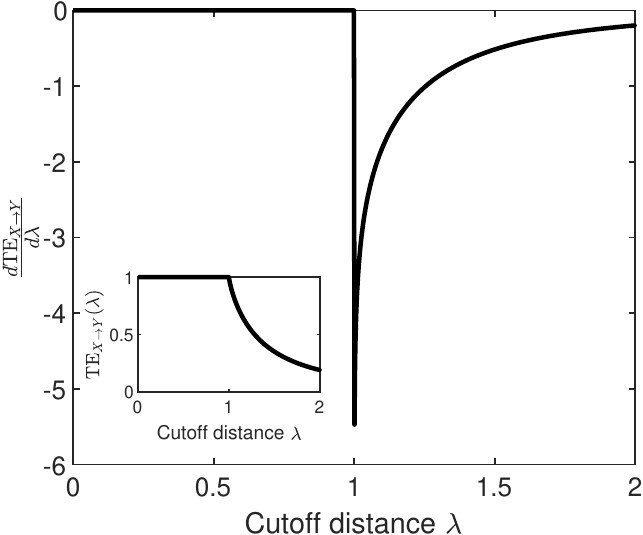}
	\caption{$\text{TE}_{X\to Y}(\lambda)$  and $\frac{d\text{TE}_{X\to Y}}{d\lambda}$ of the simplest binary model with $L=2$, $R=1$.  }
	\label{fig: binarized_TE}
\end{figure}
\begin{figure*}[hbt]
	\centering
	\includegraphics[width=0.9\linewidth]{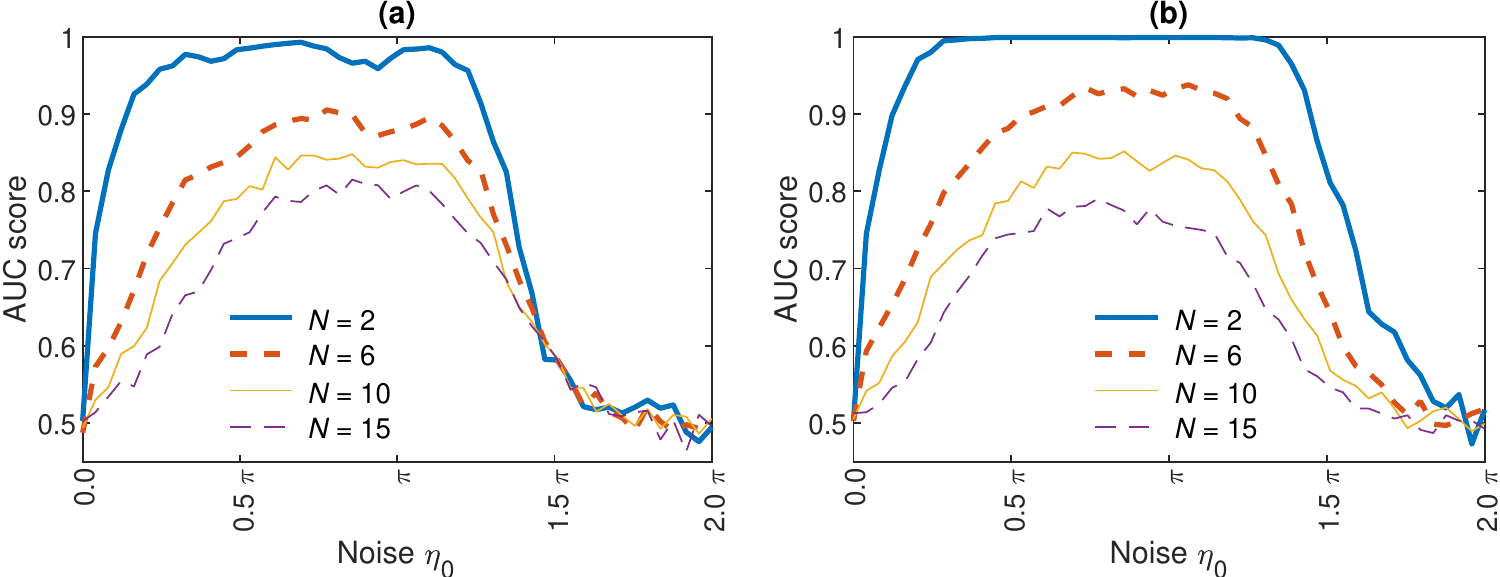}
	\caption{Classification scores of different numbers of particles at different noise levels. (a) Fixed box size. Density changes with different number of particles: $\rho=$ 0.02 arb. units ($N=2$), 0.06  ($N=6$), 0.10  ($N=10$), and 0.15  ($N=15$). (b) Fixed density as $0.1$ arb. units. Box size changes with different number of particles: $L=4.47$ arb. units ($N=2$), 7.75 ($N=6$), 10  ($N=10$), and 12.25  ($N=15$). }
	\label{fig: vary nbird vs noise}
\end{figure*}

\begin{figure}[htb]
	\centering
	\includegraphics[width=.9\linewidth]{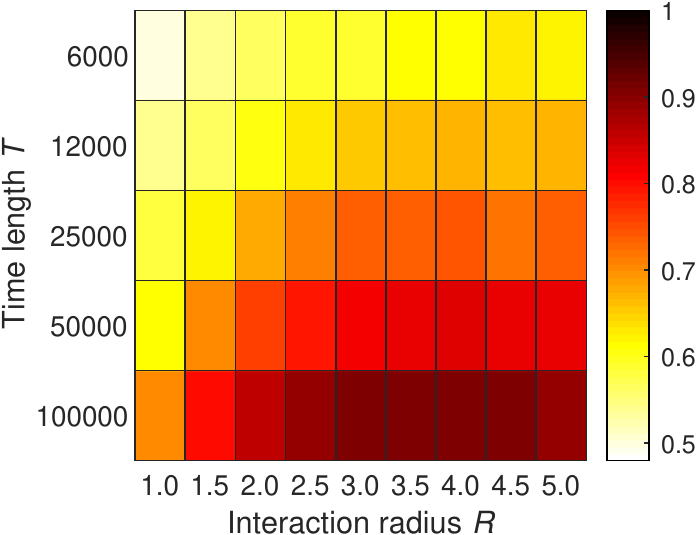}
	\caption{AUC landscape at different time length for different interaction radius. AUC landscape corresponds to $\eta_0=1.2 \pi$ arb. units and $w_{\rm LF}=1.05$ arb. units. AUC value increases as the time length $T$ and interaction radius $R$ increase. Similar behaviors were
		observed at other noise levels (not shown).}  \par
	\label{fig: vary_data_length_R}
\end{figure}

In the VM each particle moves with a constant speed $v_0$. In this paper the value of $v_0$ is set to be 0.3 arb. units. It was found that in the range of $0.05 \le v_0\leq 0.9$ the classification scores of leader and follower were almost the same. Though the speed is the same for all particles, particles change their direction of motion over time. In case of a leader-follower pair of particles, because of the domination of the leader, the follower particle changes its direction towards the leader\textquotesingle s headings.  Hence orientation of particles is used to compute transferred information between them.  
In this paper, the orientation space $[0:2\pi]$ is discretized into six bins of equal size which are represented by unique symbols \cite{PhysRevE.102.012404}. The orientation of each particle changes over time, which may produce different symbols to its time series. Finally, these sequences of symbols are used to compute TE between particles.  \par

In Eq. (\ref{eqn: weight}) the orientation of the particle $i$ at time $(t+1)$ depends largely on the orientation of itself and nearby particles at time $t$. The delay time $\tau$ is set to be 1 for the estimation of TE throughout this paper. All analyses were performed with 1000 realizations and  for each realization the values of $\vec{r}_i^t$ and $\theta_i(t)$ at time $t=0$ are chosen randomly in Eqs. (\ref{eqn: position}) and (\ref{eqn: weight}), respectively. In this paper box size $L$, number of particles $N$, time length $T$, interaction radius $R$, are varied to check their effect on classification score. \par

\section{Cutoff distance}
\label{sec: cutoff}
Knowledge of the interaction domain greatly improves the classification of leaders and followers. In practice, however, one does not know the interaction domain for a group of animals, cells, or birds a priori. To deduce it from an ensemble of trajectories, the `cutoff distance variable' $\lambda$ was introduced \cite{PhysRevE.102.012404}. In this problem setting, the interaction domain is considered as a circle of radius $R$, which is unknown, however, in general the same technique can be applied to infer an interaction domain of any shape.
Then for the estimation of TE, the cutoff distance $\lambda$ is defined as a distance up to which the interactions between particles are taken into consideration [Fig. \ref{fig: cutoff_cartoon}]. In other words, for the estimation of TE from the `symbolic time series' of a particle to another, the probability distributions are estimated only at the time instance $t$ when the distance between those two particles is less than the cutoff distance $\lambda$. For a fixed cutoff distance $\lambda$, TE from a particle $X$ to another particle $Y$ has the following form: 

\begin{equation}
\label{eq: TE_vs_lambda}
\begin{split}
\text{TE}_{X\to Y}(\lambda)&=\sum_{y_{t+\tau},y_t,x_t}p(y_{t+\tau},y_{t},x_t|d \le\lambda) \times\\
&\log_2 (\frac{p(y_{t+\tau}|y_t,x_t,d \le\lambda)}{p(y_{t+\tau}|y_t,d \le\lambda)})
\end{split}
\end{equation}
\noindent where $x_t$, $y_t$, and $d=|\vec{r}_X^t-\vec{r}_Y^t|$ represent the orientation of the particles $X$ and $Y$ at time $t$, and the distance between the particles $X$ and $Y$ at time $t$, respectively. \par
Finally the value of $\lambda$ is varied and TE between particles is computed as a function of $\lambda$. Whenever there is no mention of a cutoff distance $\lambda$, e.g. in Section V, it means that the distance information between particles is not considered which is the conventional way of TE computation \cite{PhysRevE.93.042411, orange2015transfer}.

In a group of particles, the motion of a particle is influenced by other particles lying within its interaction domain. When two particles enter into their interaction domain, they share or transfer information, resulting in some change in their movements. Such information flow should decrease as the distance between the two becomes greater than the interaction domain, and goes to zero at the limit of the distance goes to infinity because of loss of interactions. For the transfer entropy (Eq. \ref{eq: TE_vs_lambda}) where all pairs satisfying $d \le\lambda$ are taken into account, when the cutoff distance $\lambda$ exceeds the underlying interaction domain $R$, $\text{TE}_{X\to Y}(\lambda)$ should decrease because it takes into account not only interacting pairs of particles but also non-interacting, independently-moving pairs, which should ``dilute'' the information flow among interacting pairs in the system.
The magnitude of negative gradient of the transfer entropy with respect to $\lambda$ after exceeding $R$ is expected not to be a constant along $\lambda$ but the magnitude becomes get smaller as $\lambda$ gets larger. When the two particles are within interaction domain, more or less their motility is influenced to each other so that the transfer entropy is expected to have some finite value with some fluctuation for $\lambda \le R$.

We now derive a simplest analytic model to manifest the relationship between the interaction domain $R$ and transfer entropy $\text{TE}_{X\to Y}(\lambda)$ as follows: 
Consider an agent $X$ represented as a binary random process, 
i.e. $X_{t}=0$ with probability $\frac{1}{2}$ and $X_{t}=1$ with probability $\frac{1}{2}$ (similar to ``fair coin toss''), whose value influences to another agent's future value $Y_{t+1}$ depending on an auxiliary variable $Z_t$. $Z_{t}$ is a continuous, uniformly distributed variable on the interval $[0, L]$, and 
$Y_{t+1} = X_{t}$ whenever $Z_{t} \le R$, otherwise $Y_{t+1}$ is a binary random process as $X$, where $Y(0)$ = 0.  Namely,
$R (\le L)$ serves a similar function as the interaction radius in the VM model, and $Z$ does
as the distance between a pair of particles.  In a more extreme sense of the VM, $Y_{t+1}$ depends deterministically on $X_{t}$ whenever $Z_t$ below $R$ and $Y_{t+1}$ is completely random otherwise. Therefore intuitively, the function $\text{TE}_{X\to Y}(\lambda)$ should have a similar shape to that of the VM for the binary system, especially near the value $\lambda=R$. The benefit of the binary system is that we can compute  $\text{TE}_{X\to Y}(\lambda)$ analytically. 
$\text{TE}_{X\to Y}(\lambda)$ is given as follows:
\begin{equation}
\label{eq: analytical_TE}
\text{TE}_{X\to Y}(\lambda)=\frac{\lambda+R}{2\lambda} \log_2\frac{\lambda+R}{\lambda}+\frac{\lambda-R}{2\lambda} \log_2\frac{\lambda-R}{\lambda}
\end{equation}
whenever $\lambda > R$ (see Appendix C for its derivation). When $\lambda \le R$,  $\text{TE}_{X\to Y}(\lambda)$ is equal to 1 since the dynamics for $Y_{t+1}$ $Z_t \le R$ is always deterministically dependent on $X_t$.  The derivative $\frac{d\text{TE}_{X\to Y}}{d\lambda}$ when $\lambda > R$ is given by 
\begin{equation}
\label{eq: analytical_dTE}
\frac{d\text{TE}_{X\to Y}}{d\lambda}=\frac{R}{2\lambda^2}\rm{log}_2\frac{\lambda-R}{\lambda+R}.
\end{equation}
The values of $\text{TE}_{X\to Y}(\lambda)$  and $\frac{d\text{TE}_{X\to Y}}{d\lambda}$ for the parameters $R=1$ and $L=2R$ are shown in  Fig.~\ref{fig: binarized_TE}.
The function ${\text{TE}}_{X\to Y}(\lambda)$ has a clear visible kink at $\lambda=R$, which is made more apparent by looking at the derivative function $\frac{d \text {TE}_{X\to Y}(\lambda)}{d \lambda}$ of the inset in Fig. \ref{fig: binarized_TE}. By computing  $\text{TE}_{X\to Y}(\lambda)$ from trajectory data and adequately detecting this kink, one may readily infer the interaction domain without any prior knowledge about the system, as we will demonstrate in the following section.

\section{Results and discussions}

\begin{figure}[hbt]
	\centering
	\includegraphics[width=0.9\linewidth]{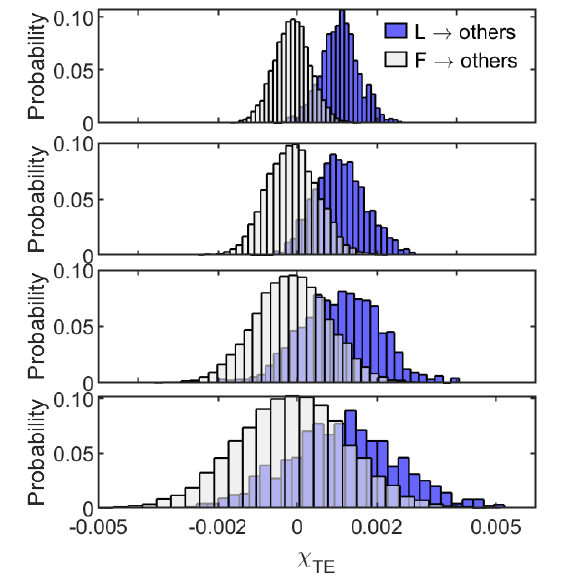}
	\caption{ Distributions of the classifiers $\chi_{\rm TE}$(/bits) from the leader
		to the others, and those from a follower to the others for $R=3$, $\eta_0=1.2\pi$, and (a) $T=100,000$ arb. units, (b) $T=50,000$, (c) $T=25,000$, and (d) $T=12,000$.}
	\label{fig: distribution vary length fixed R}
\end{figure}

\begin{figure}[hbt]
	\centering
	\includegraphics[width=0.9\linewidth]{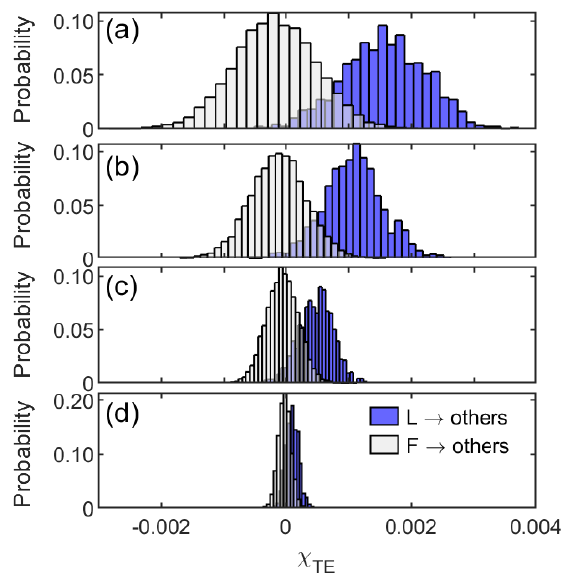}
	\caption{  Distributions of the classifiers $\chi_{\rm TE}$(/bits) from the leader
		to the others, and those from a follower to the others for $T=100,000$, $\eta_0=1.2\pi$, and (a) $R=4.0$, (b) $R=3.0$, (c) $R=2.0$, and (d) $R=1.0$.}
	\label{fig: distribution fixed length vary R}
\end{figure}
\begin{figure}[htb]
	\centering
	\includegraphics[width=.9\linewidth]{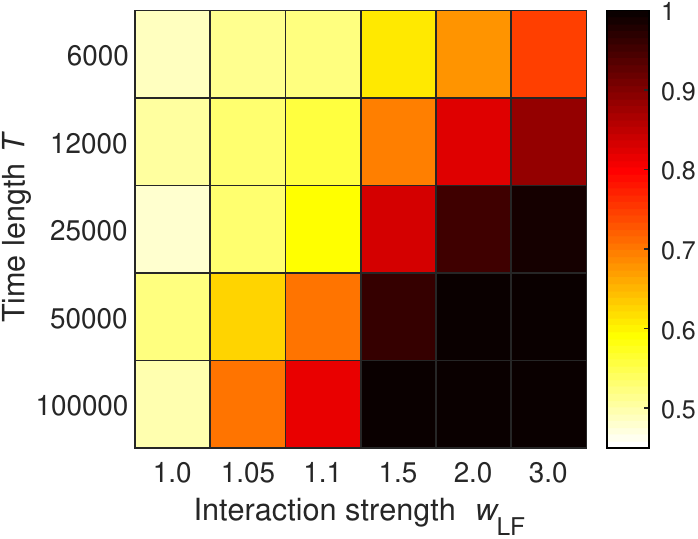}
	\caption{AUC landscape at different time length for different interaction strength. AUC landscape corresponds to $\eta_0=1.2 \pi$ arb. units and $R=1$ arb. units. For this analysis $w_{\rm FL}=1$ is fixed and $w_{\rm LF}$ were varied. AUC value increases as the time length $T$ and  leaders\textquotesingle ~ interaction strength $w_{\rm LF}$ increase. Similar behaviors were
		observed at other noise levels (not shown).}   \par
	\label{fig: vary_data_length_w}
\end{figure}
   
\begin{figure*}
	\centering
	\includegraphics[width=.9\linewidth]{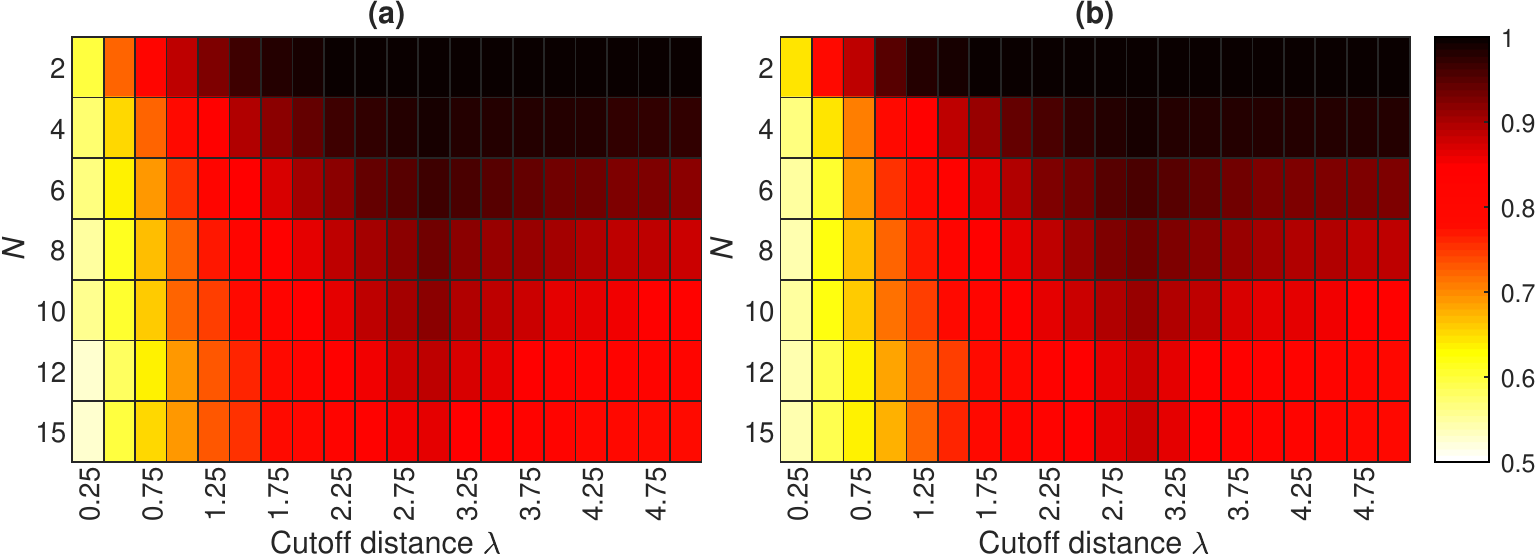}
	\caption {AUC landscape with respect to number of particles and cutoff distance. Actual interaction radius $R$ used for the trajectory calculation is $3.0$ and the noise is  $\eta_0=\frac{\pi}{2}$. (a) Fixed box size. The highest AUC scores and their locations are: 0.999 at $\lambda=3.0$ arb. units $(N=2)$, 0.987 at $\lambda=3.0$ $(N=4)$, 0.963 at $\lambda=3.0$ $(N=6)$, 0.935 at $\lambda=3.0$ $(N=8)$, 0.915 at $\lambda=3.0$ $(N=10)$, 0.890 at $\lambda=3.0$ $(N=12)$, and 0.867 at $\lambda=3.0$ $(N=15)$. (b) Fixed density. The highest AUC scores and their locations are: 0.999 at $\lambda=3.0$ $(N=2)$, 0.987 at $\lambda=3.0$ $(N=4)$, 0.959 at $\lambda=3.0$ $(N=6)$, 0.937 at $\lambda=3.0$ $(N=8)$, 0.910 at $\lambda=3.0$ $(N=10)$, 0.880 at $\lambda=3.0$ $(N=12)$, and 0.877 at $\lambda=3.0$ $(N=15)$.  }
	\label{fig: cutoff_nbird}
\end{figure*}
\begin{figure*}[htp]
	\centering
	\begin{minipage}{1\linewidth}
		\includegraphics[width=\linewidth]{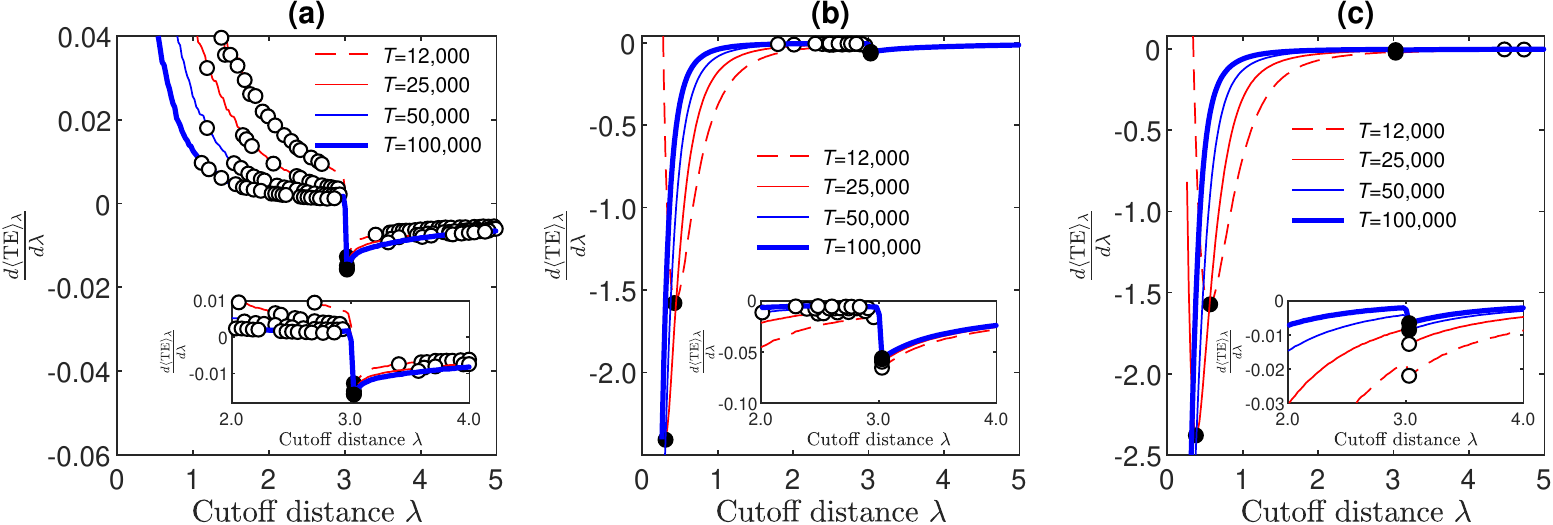}
	\end{minipage}{\vspace{3mm} }
\caption{Derivative of average TE for different time length along with local minima and identified interaction radius based on global minimum scheme at (a) $\eta_0=0.2\pi$, (b) $\eta_0=1.2 \pi$ and (c) $\eta_0=1.8\pi$. The actual interaction radius $R$ used for ensemble of trajectories is 3. Circles represent all local minima and filled-circles represent the identified interaction radius at each time length $T$. } 
\label{Fig. change point minima}
\end{figure*}

\begin{figure*}[htp]	
	\begin{minipage}{1\linewidth}
		\includegraphics[width=\linewidth]{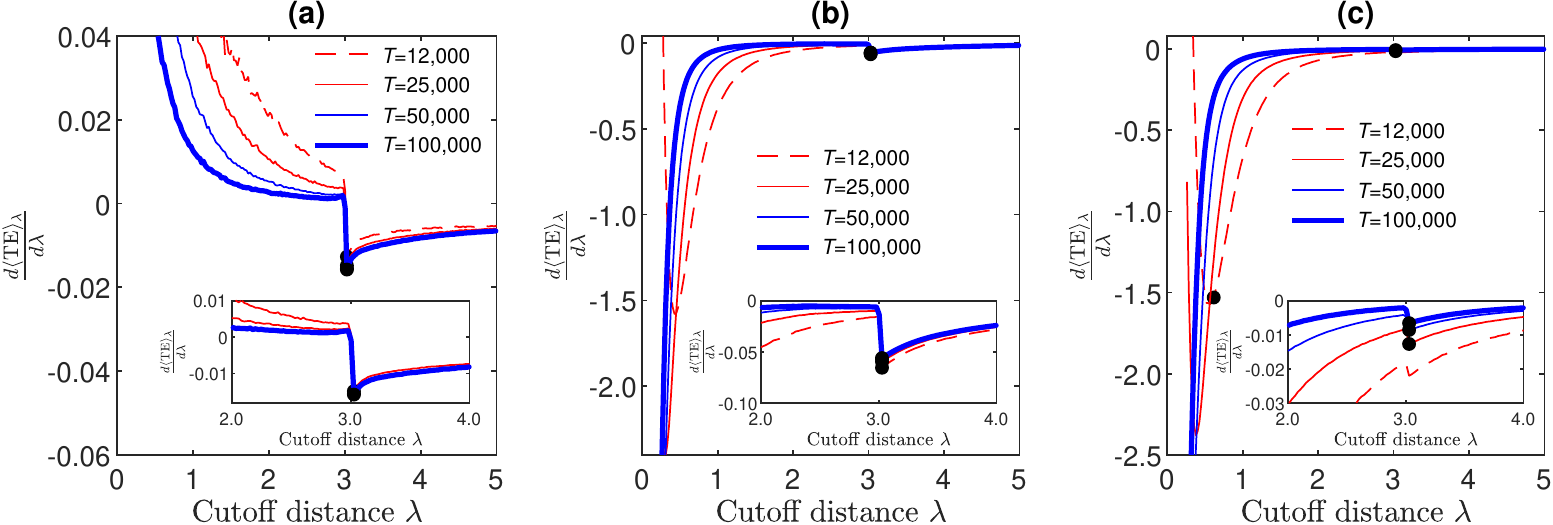}
	\end{minipage}
	\caption{Derivative of average TE for different time  lengths along with the identified interaction radius based on convexity score scheme at (a) $\eta_0=0.2\pi$, (b) $\eta_0=1.2 \pi$ and (c) $\eta_0=1.8\pi$ for $\{M\} =\{M| 2\le M \le 30\}$ and $\delta=1\times10^{-4}$.}
	\label{Fig: change point cp}
\end{figure*}

Figure \ref{fig: vary nbird vs noise} shows the AUC classification score (=the extent of how leader and followers are correctly classified from their orientation dynamics) with different numbers of particles $N$ and noise levels $\eta_0$. Here, only one particle serves as a leader and the other ($N-1$) particles serve as followers, under the constraint of box size $L$ [Fig. \ref{fig: vary nbird vs noise}(a)] or density $\rho$ [Fig. \ref{fig: vary nbird vs noise}(b)]. At  low noise level $\eta_0 \sim 0$ and very high level $\eta_0 > 1.5\pi$, the distributions of transfer entropies $\chi_{\rm TE}$ from leader to follower and vice versa over 1000 realizations were found to significantly overlap with each other, which makes differentiation between leader and follower difficult. The low AUC at very low noise level arises from the fact that particles fall into some concerted motion very quickly dependent solely on initial configurations, resulting in insufficient sampling of orientational dynamics,  and in turn the low AUC at very high noise arises from overshadowing of the interactions by random noises \cite{PhysRevE.102.012404}.
As the number of particles increases, the AUC score decreases in both fixed box size and fixed density cases. This is because, in any pair of particles for which TE is evaluated, their motions are also influenced by the other particles, and the more the number of particles increases, the more the motions of the particles in question are influenced by the third (or higher) particle.  \par

The effects of time length $T$, and interaction radius $R$ on classification score are shown in Fig. \ref{fig: vary_data_length_R}. In this analysis 10 particles were used, with one serving as a leader and the other 9 particles as followers. In Fig. \ref{fig: vary_data_length_R} we varied time length $T$ and interaction radius $R$. It is shown that the AUC score increases with $T$ and $R$. For shorter $T$, due to insufficient sampling in characterizing leader-follower interaction relationship, the distributions of $\chi_{\rm TE}$ from leader to the others and from follower to the others have higher variance as shown in Fig. \ref{fig: distribution vary length fixed R}. As a result, it is difficult to distinguish leader and followers for short $T$. As $T$ increases, due to more exploration of interaction events between leader and followers, the variance of leaders' and followers' distributions gets smaller, making the classification easier. Figure \ref{fig: distribution fixed length vary R} shows, in turn, the interaction radius $R$ dependency on distributions of the classifiers $\chi_{\rm TE}$. Larger $R$ allows particles to be taken into account in elucidating the leader-follower interaction relationship, which produces easily distinguishable distributions of leader and followers as shown in Fig. \ref{fig: distribution fixed length vary R}(a). In contrast, when $R$ is smaller, the $\chi_{\rm TE}$ distributions of leaders and followers overlap each other more with smaller variance,  making the classification more difficult. \par

In Fig. \ref{fig: vary_data_length_w}  we set $R=1.0$ and varied time length and interaction strength of leader on follower $w_{\rm LF}$. In this analysis followers' interaction strength is set to 1.0, i.e.,  $w_{\rm FL}=w_{\rm FF}=1.0$. Hence $w_{\rm LF}=1.0$ represents no leader case that produces AUC close to 0.5 as expected. As the $w_{\rm LF}$ increases AUC value also increases as the leader is getting more influential on followers which makes classification easier even at short time length. \par

How does the predictability of interaction radius by using TE with cutoff distance depend on the number of particles? Figure \ref{fig: cutoff_nbird}
represents the AUC landscape as a function of number of particles $N$ and  cutoff distance $\lambda$. Figure \ref{fig: cutoff_nbird}(a) corresponds to a fixed box size (i.e. density is changing with $N$), whereas Fig. \ref{fig: cutoff_nbird}(b) represents the systems having same density (i.e. $L$ is changing with $N$). The actual interaction radius used to simulate the trajectories of the particles was $R=3$ and noise was set to $\eta_0=\frac{\pi}{2}$. Although at each $N$ the maximum AUCs are located at the underlying interaction radius $R=3.0$, and the maximum AUC score is higher than the conventional no-cutoff scheme corresponding to $R=5\sqrt2$,  the maximum AUC decreases as the number of particles increases due to the increase of indirect interactions between particles. Similar behaviors are observed at different interaction radius $R$ and noise levels (not shown here). \par

\begin{figure*}[htp]
	\centering
	\begin{minipage}{1\linewidth}
		\includegraphics[width=\linewidth]{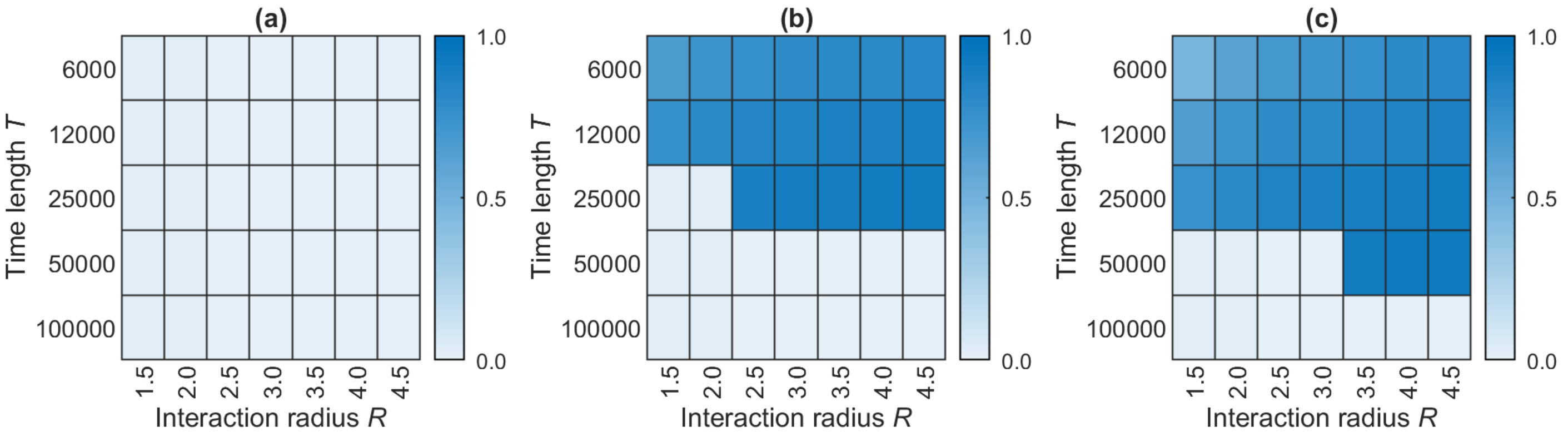}
	\end{minipage}
	\caption{ Relative error $\hat{R}$ in identifying underlying interaction domain using global minimum scheme at (a) $\eta_0=0.2\pi$, (b) $\eta_0=1.2\pi$, and (c) $\eta_0=1.8\pi$. }
	\label{Fig: relative error min}
\end{figure*}

	\begin{figure*}[htp]
		\centering\begin{minipage}{1\linewidth}
			\includegraphics[width=\linewidth]{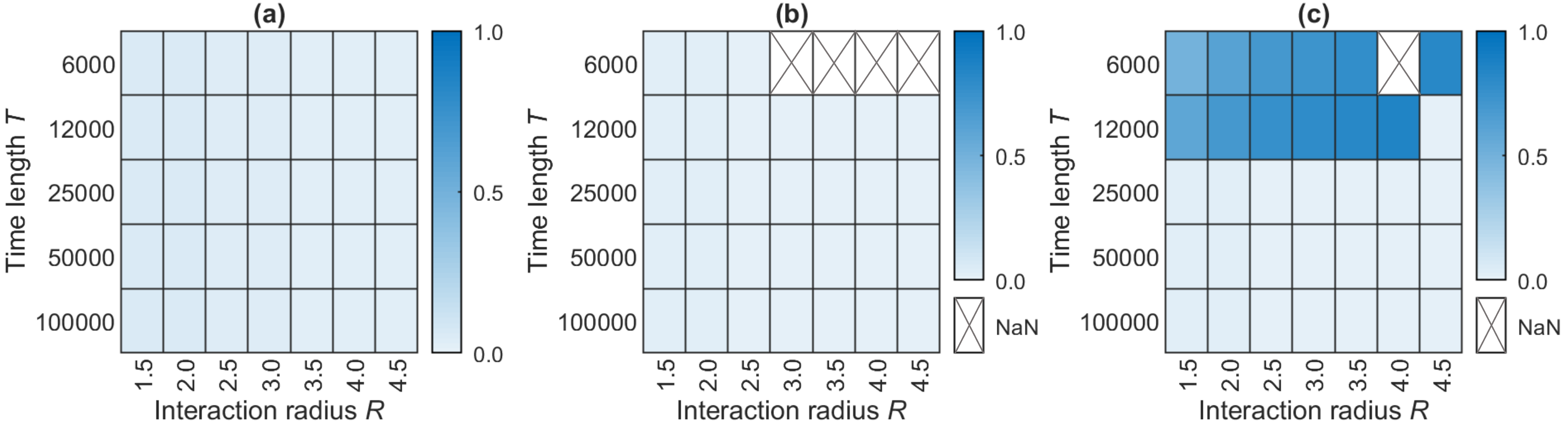}
		\end{minipage}
		\caption{ Relative error $\hat{R}$ in identifying underlying interaction domain using convexity score scheme with $\delta=1\times 10^{-4}$ and  $\{M\} =\{M| 2\le M \le 30\}$   at (a) $\eta_0=0.2\pi$, (b) $\eta_0=1.2\pi$, and (c) $\eta_0=1.8\pi$.  Cross-marked boxes `NaN' mean that the scheme fails to identify the interaction radius.} 
	\label{Fig: relative error cp}
\end{figure*}

How can one infer the underlying interaction radius solely from ensembles of trajectories? Recently, a simple scheme has been proposed  \cite{PhysRevE.102.012404} to infer the underlying interaction distance from ensembles of trajectories, based on the existence of a significant decrease in averaged transfer entropy when cutoff distance $\lambda$ exceeds the underlying interaction radius. The average TE for a specific cutoff distance $\lambda$, $\langle \text{TE} \rangle_\lambda$ has the following form:
\begin{equation}
\langle \text{TE} \rangle_\lambda=\frac{1}{s}\sum_{k=1}^{s}\left[\frac{1}{N(N-1)}\sum_i\sum_j\text{TE}_{i\to j}^{(k)}(\lambda)\right]
\end{equation}
where $s$ represents the number of realizations (generating from different initial conditions) for which we used 1000,  and $N$ is the number of particles in the system. Here $\text{TE}_{i\to j}^{(k)}(\lambda)$ represents the TE from the particle $i$ to the particle $j$ for cutoff distance $\lambda$ at $k\rm{th}$ realization. \par
The interaction distance is inferred as that where the minimum of the derivatives exists along the cutoff distance $\lambda$:
\begin{equation*}
\label{eqn: min_derivative}
\hat{R} \equiv {\rm argmin}_{\lambda} \frac{d \langle {\text {TE}} \rangle_\lambda}{d \lambda},
\end{equation*} 
under the condition of  $\frac{d^2 \langle {\rm TE} \rangle_\lambda}{d \lambda ^2}=0$.  
In practice, the length of trajectories may not be long enough and shorter length tends to result in a fluctuation in the course of TE along the cutoff distance $\lambda$, resulting in apparent minima.\par

Figure \ref{Fig. change point minima} shows the derivative of  average TE as a function of cutoff distance $\lambda$, denoted by $\frac{d\langle  \text{TE} \rangle_\lambda }{d\lambda}$ for  interaction radius $R=3$ at (a) $\eta_0=0.2\pi$, (b) $\eta_0=1.2\pi$, and (c) $\eta_0=1.8\pi$ for different $T$. Here circles represent all local minima and the filled-circles represent the global minimum identified as the interaction radius.  
In Figs. \ref{Fig. change point minima}(b) and \ref{Fig. change point minima}(c) for relatively short $T=25,000$ or less,   $\frac{d\langle  TE \rangle_\lambda }{d\lambda}$ as a function of $\lambda$ has the global minimum at low $\lambda$.
The locations of these global minima for short time length change with $T$ and they vanish when $T$ is longer, and the longer $T=50,000-100,000$ both result in a close value of the underlying interaction radius $R=3$. 
This implies that to look for global minimum of derivative of transfer entropy may not necessarily result in an approximation of the true interaction radius especially for some short $T$ at high noise levels. \par

In this paper, we present another scheme expected to be robust against fluctuations of average transfer entropies along the cutoff distance by introducing a measure to quantify the degree of strong convexity at coarse-grained level, and time variance of underlying interaction radius of particles as follows. 

 
Note that in Fig. \ref{fig: binarized_TE} that $\frac{d \text {TE}_{X\to Y}(\lambda)}{d \lambda}$ has a discontinuity at the position of $\lambda=R$.
This is because the function $\text{TE}_{X\to Y}(\lambda)$ is unchanging when $\lambda \le R$, since the dynamics for that system are unchanging for that interval. However as $\lambda$ increases above $R$, portions of time series where $X$ and $Y$ are not interacting begin to be included, and $\text{TE}_{X\to Y}(\lambda)$  begins to drop. This change from a zero derivative to a negative derivative is abrupt, and thus a discontinuity in $\frac{d \text {TE}_{X\to Y}(\lambda)}{d \lambda}$ is observed. This change can be detected by either a minimum in the derivative or a maximum in convexity, however, as we will elucidate further for the VM, the local minimum technique fails when the length of trajectories are short. For the VM,  in the insets of Fig. \ref{Fig. change point minima} the shape of 
$\frac{d\langle  \text{TE} \rangle_\lambda }{d\lambda}$ as a function of $\lambda$ is (strongly) convex  near the actual interaction radius irrespective of time length $T$, while spurious local minima tend to appear at short cutoff distances. 
This indicates that the derivative of transfer entropy as a function of cutoff distance can shed light on the underlying spatial scale of interactions among particles. However, it is not trivial to devise a scheme to automatically infer the interaction radius. Since $\frac{d\langle  \text{TE} \rangle_\lambda }{d\lambda}$ as a function of $\lambda$ is convex  near the interaction radius, 
a measure of convexity  of  $\frac{d\langle  \text{TE} \rangle_\lambda }{d\lambda}$  is versatile  in determining the interaction radius. In general, due to noise, $\frac{d\langle  \text{TE} \rangle_\lambda }{d\lambda}$ can be fluctuated, producing apparent convex patterns locally. Thus in defining the convexity score, it is necessary to capture the non-local feature of $\frac{d\langle  \text{TE} \rangle_\lambda }{d\lambda}$ rather than the local feature that may be subject to noise(s). 
We define the convexity score $\kappa(\lambda_i)$ of a function $f(\lambda)$ at a point $\lambda_i$ as $\kappa(\lambda_i)=\frac{1}{M}\sum_{m=1}^{M}\sigma_{i}(m)$ where $\sigma_{i}(m)=+1$ if $f(\lambda_{i-m})-f(\lambda_i)>\delta$ and $f(\lambda_{i+m})-f(\lambda_i)>\delta$ and $\sigma_{i}(m)=-1$ if $f(\lambda_i)- f(\lambda_{i-m})>\delta$ and $f(\lambda_i)- f(\lambda_{i+m})>\delta$, otherwise $\sigma_{i}(m)=0$. Here $\delta$ is a non-negative small number and $M$ is the number of neighboring points used to determine the convexity score,
and $-1 \le \kappa(\lambda_i) \le 1$. Here the function $f(\lambda)$ represents the derivative of average TE, $\frac{d\langle  \text{TE} \rangle_\lambda }{d\lambda}$.


Thus, around a point $\lambda_i$ where $\frac{d\langle  \text{TE} \rangle_\lambda }{d\lambda}$ is  convex at some coarse-grained level 
$\kappa(i)$ is close to unity. Hence the point $\lambda_i$ around which  $\kappa(\lambda_i)$ is the maximum is considered to be an indicator of the interaction radius above which average transfer entropy between particles significantly decreases. 

How to choose the optimal $M$ and $\delta$? 
We define the estimated interaction radius $\hat{R}$ as $\hat{R} \equiv {\rm argmax}_{\lambda} \kappa(M;T)$, where $T$ represents the time length. Then the cost function is defined as
\begin{equation}
\label{eqn: cost}
C(M) \equiv \sum_T\sum_{T'}|\hat{R}(M;T)-\hat{R}(M;T')|,
\end{equation}
  by assuming that the interaction radius is independent of time, i.e., time-invariant, and there exists (approximately) sufficient statistics for each time length in elucidating the interaction events. The parameter $M$ is determined so that $M={\rm argmin}_{M \in \{M\}} C(M)$. Here the set of $M$ to be searched for finding optimal $M$, $\{M\}$, is this users need to input {\it a priori}.  Note that for some time length $T$, $\hat{R}(M;T)$ could not be chosen uniquely due to the degeneracy of 
$\kappa (M;T)$ and also $\hat{R}(M;T)$ could become undefined when $\frac{d\langle  \text{TE} \rangle_\lambda }{d\lambda}$ curve does not  has any strongly convex part. In both the cases, we exclude such $T$ in Eq. (\ref{eqn: cost}) to compute the cost function $C(M)$ in defining optimal $M$. We also varied  $\delta=1\times 10^{-8},1\times 10^{-6},1\times 10^{-4}$ (arb. units) and found that within these ranges of $\delta$ has no significant effect on the estimation of interaction radius.   


Figures \ref{Fig: relative error min} and \ref{Fig: relative error cp} show the relative errors ($\Delta R$) in identifying the underlying interaction domain using the global minimum of the derivative of transfer entropy and convexity score scheme, respectively, as a function of time length and interaction radius at three different noise levels.
Here relative error ($\Delta R$) is defined as $\Delta R=\frac{|R-\hat{R}|}{R}$, where 
$\hat{R}$ is the identified interaction radius using either of the two schemes. \par 

For the global minimum (of TE derivative) scheme [Fig.~\ref{Fig: relative error min}], there exists a clear trend such that the larger the noise level the larger the relative error, and as the time length $T$ decreases, the relative errors are more pronounced for relatively large noise levels  $\eta_0=1.2\pi-1.8\pi$. 
Because of the appearance of a minimum at low cutoff distance for short $T$ that ceases to exist for longer $T$  in Figs.~\ref{Fig. change point minima}(b) and \ref{Fig. change point minima}(c), the global minimum scheme apparently possesses higher relative error for short $T$  [Figs. \ref{Fig: relative error min} (b) and  \ref{Fig: relative error min} (c)]. 


Figure \ref{Fig: relative error cp} shows the relative error in identifying the interaction radius at different noise levels using convexity score scheme. Although global minimum scheme possesses high relative error at moderate noise when data length is short [Fig. \ref{Fig: relative error min}(b)] but the convexity score scheme 
identifies the interaction radius satisfactorily  [Fig. \ref{Fig: relative error cp}(b)] 
for $T\ge12,000$. But for short $T$, e.g., $T\le6000$, the convexity score scheme fails to identify the interaction radius for $R\ge3$.   
Like the global minimum scheme, the convexity score scheme possesses high relative error $\Delta R$ when noise level is very high [Fig. \ref{Fig: relative error cp}(c)] and $T$ is short ($T\le 12,000$). But when $T$ is large ($T\ge$ 25,000), the convexity score scheme can identify the interaction radius competently [Fig. \ref{Fig: relative error cp}(c)]  even at high noise levels. 


\section{Conclusions}

 In this study, we examined the performance of two heuristic schemes using the derivative of transfer entropy with respect to cutoff distance $\lambda$, $\frac{d\langle \text{TE} \rangle_\lambda }{d\lambda}$: global minimum and convexity score scheme for determining the interaction radius by using the modified VM. The striking feature 
 -based on which we proposed the two schemes- is that $\frac{d\langle \text{TE} \rangle_\lambda }{d\lambda}$ exhibits a kink near the actual interaction radius. A method that is capable of determining the exact location of that kink can be used in inferring interaction radius.  
  
 For short time length (at the moderate and high level of noise) for the modified VM, the derivative of average TE exhibits a minimum at low cutoff distances that produce a relatively high error for the global minimum scheme. Moreover, in real experiments 
 it is not necessarily possible to get sufficiently long trajectories with small noises, 
 and the global minimum scheme may yield some non-negligible error especially for short trajectories with noise. 
In this paper, an alternative scheme was presented, based on the property of convexity of  $\frac{d\langle \text{TE} \rangle_\lambda }{d\lambda}$ at the coarse-grained level and the assumption of time-invariance of the underlying interaction domain.
For the modified VM, the scheme could capture the underlying interaction radius at the low and moderate levels of noise especially for relatively short time length, ca. $T$=12,000-25,000 for which global minimum scheme possesses high relative error (at moderate noise level). 
These two heuristic schemes are compliment to each other and, as for appropriate usages, one should first visualize transfer entropy as a function of cutoff distance $\lambda$ with its derivatives with respect to cutoff distance to confirm the existence of abrupt changes along the cutoff distance.
Significant changes in the derivative of TE with respect to cutoff distance were also observed for classical trajectories of particles interacting via Lennard-Jones potential (not shown), and the existence of some significant change along cutoff distance around the typical length scale of interactions may be ubiquitous.

 In systems with many variables, identifying causal relationships is a daunting task. An important aspect of systems that should be exploited, however, is that a particular variable may be only influencing another particular variable at certain time instances. We have shown that filtering out those time instances where influence does not occur greatly improves the identification of causal relationships. In the Vicsek model, for example, two agents may only interact when their distance is less than a certain threshold. To pose this as a question, we wonder at which value of interaction radius $R$ does the motion of one agent influence the motion of another? More generally, we ponder the question: at which levels of variable $X$ does variable $Y$ influence variable $Z$?  In future work, we 
 will demonstrate the applicability of this method by applying it to data sets stemming from different fields. 

\par

\begin{acknowledgements}
	 This work was supported by a Grant-in-Aid for Scientific Research on Innovative Areas ``Singularity Biology (No.8007)'' (18H05413), MEXT, the research program of `Five star Alliance' in `NJRC Matter and Dev' (No. 20191062-01), and by JSPS (No. 25287105 and 25650044 to T.K.), and JST/CREST (No. JPMJCR1662 to T.K.).
\end{acknowledgements}

\section*{DATA AVAILABILITY}
The data that support the findings of this study are available from the corresponding author upon request.
      \appendix

      \section{Estimated interaction radius at $R=2.0$}
\begin{figure*}[htp]	
      	\begin{minipage}{1\linewidth}
      		\includegraphics[width=\linewidth]{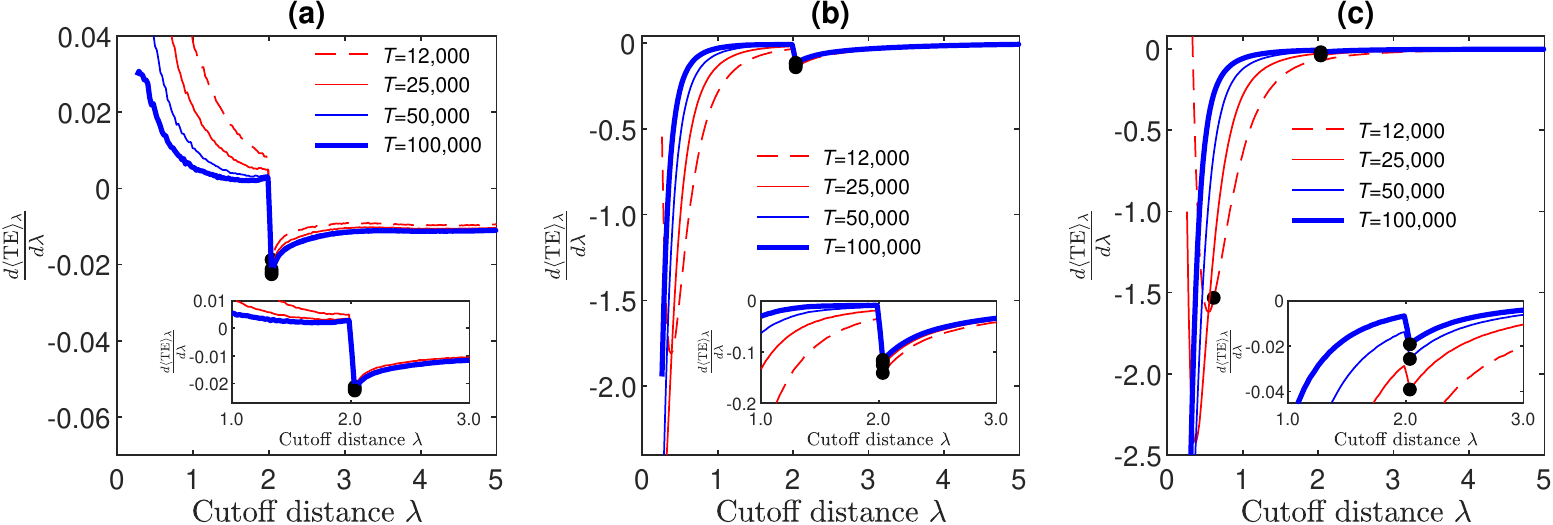}
      	\end{minipage}
      	\caption{Derivative of average TE for different time  lengths along with the identified interaction radius based on convexity score scheme at (a) $\eta_0=0.2\pi$, (b) $\eta_0=1.2, \pi$ and (c) $\eta_0=1.8\pi$ for $R=2.0$, $\delta=1\times10^{-4}$ and $\{M\} =\{M| 2\le M \le 30\}$.}
      	\label{Fig: change point cp_R_2}
      \end{figure*}

      Figure (\ref{Fig: change point cp_R_2}) shows the derivative of average TE for different time  lengths along with the identified interaction radius based on convexity score scheme at different noise levels. It has been found that at low and moderate noise levels the convexity score scheme identifies the interaction radius satisfactorily [Figs. \ref{Fig: change point cp_R_2}(a) and \ref{Fig: change point cp_R_2}(b)]. But at very high level of noise ($\eta_0\approx 1.8\pi$) this scheme performs satisfactorily for longer $T$ but possesses higher relative error for shorter $T$  [Fig. \ref{Fig: change point cp_R_2}(c)]. 
      \section{Convexity score}
\begin{figure*}[htp]	
      	\begin{minipage}{0.32\linewidth}
      		\includegraphics[width=\linewidth]{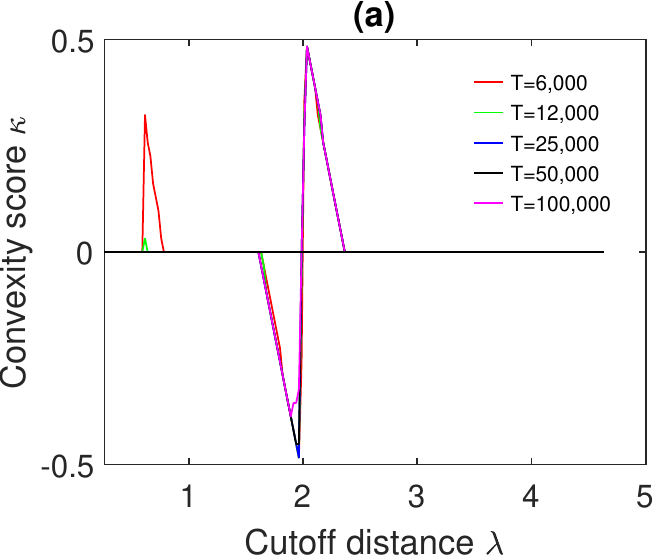}
      	\end{minipage}
      \begin{minipage}{0.32\linewidth}
      	\includegraphics[width=\linewidth]{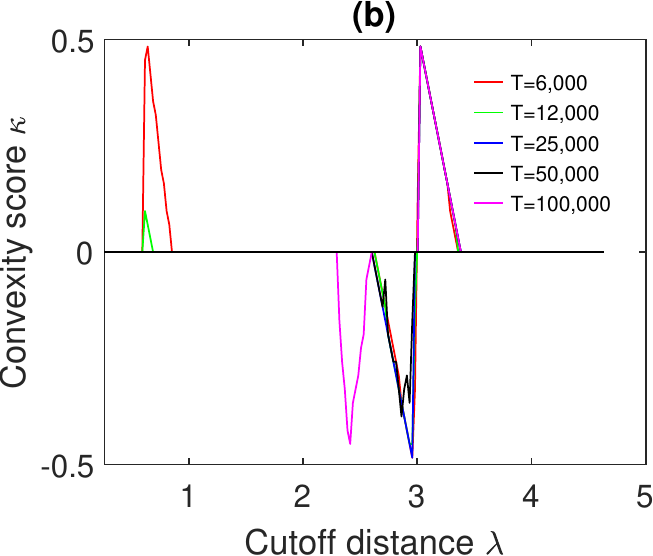}
      \end{minipage}
  \begin{minipage}{0.32\linewidth}
  	\includegraphics[width=\linewidth]{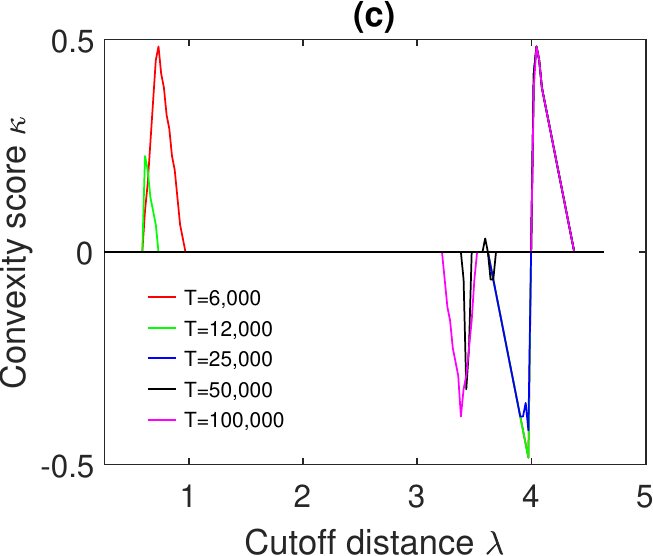}
  \end{minipage}
      	\caption{Convexity score $\kappa$ at (a) $R=2.0$, (b) $R=3.0$, and (c) $R=4.0$ at $\eta_0=1.2\pi$.}
      	\label{Fig: convexity_score}
      \end{figure*}

      According to the definition,  convexity score $\kappa$ is maximum at the point for which the   $\frac{d\langle  \text{TE} \rangle_\lambda }{d\lambda}$  is strongly convex, and $\kappa$ is minimum where the curve is concave. Figure (\ref{Fig: convexity_score}) shows the convexity score for moderate noise level ($\eta_0=1.2\pi$) at different interaction radii $R$. For $R=2.0$ [Fig. \ref{Fig: convexity_score}(a)]  unique maximum convexity scores have been identified for each $T$ ($T=6,000-100,000$). Hence the convexity score scheme identifies the interaction radius correctly. For $R=3.0$ and $R=4.0$, no unique $\lambda$ was identified
for $T=6,000$
[Figs. \ref{Fig: convexity_score}(b) and \ref{Fig: convexity_score} (c)].  However, the convexity score scheme can identify the interaction radius perfectly for longer $T$ ($T=12,000-100,000$).
      
      \section{Derivation of analytical cutoff function for binary system}
      \label{sec: binary_derivation}
      To analytically derive 
       $\text{TE}_{X\to Y}(\lambda)$ for the binary system described at the end of  Sect. \ref{sec: cutoff} 
       we must compute 
       Eq.~\ref{eq: TE_vs_lambda} for the binary system that yields 
       $\text{TE}_{X\to Y}(\lambda)$ in Eq.~\ref{eq: analytical_TE}. Using the chain rule for conditional probability and replacing $x_t$, $y_t$, and $d$ by $X_t$, $Y_t$, and $Z_t$, respectively, from our binary system, Eq.~\ref{eq: TE_vs_lambda}  can be written as  
\begin{align}
\label{eq: binary_TE_log}
\rm{TE}_{X \to Y}(\lambda)&=\sum_{Y_{t+1}}\sum_{X_{t}}\sum_{Y_{t}}P(Y_{t+1},Y_t,X_t | Z_t\le \lambda)\times \nonumber \\
& \text{log}_2 \frac{P(Y_{t+1},Y_t,X_t | Z_t\le \lambda)P(Y_t|Z_t\le \lambda)}{P(Y_{t+1},Y_t|Z_t\le \lambda)P(Y_t,X_t|Z_t\le \lambda)}.
\end{align}
$P(Y_t|Z_t\le \lambda)= P(Y_t) = \frac{1}{2}$ 
since $Y_t \perp Z_t$ (where $A \perp B$ denotes $A$ and $B$ are independent), and $P(Y_t) = \frac{1}{2}$ 
by definition.
       Since $X_t \perp Y_t$ and $Y_t \perp Y_{t+1}$, $P(Y_t,X_t|Z_t\le \lambda)=P(Y_t,X_t)=P(Y_t,Y_{t+1}|Z_t\le \lambda)=P(Y_t,Y_{t+1})=\frac{1}{4}$.
       $P(Y_{t+1},Y_t,X_t  | Z_t\le \lambda) = P(Y_{t+1},X_t| Z_t\le \lambda)P(Y_t) =\frac{1}{2}P(Y_{t+1},X_t| Z_t\le \lambda)$ since $Y_t \perp (Y_{t+1},X_t,Z_t)$  and $P(Y_t) = \frac{1}{2}$. 
      Thus, $\lambda$ dependency of $\rm{TE}_{X \to Y}(\lambda)$ arises from $P(Y_{t+1},X_t | Z_t\le \lambda)$. Now we first compute a general form of $P(Y_{t+1},X_t)$ which holds irrespective of $\lambda$ and then re-write it in terms of $\lambda$ shortly thereafter. Here $P(Y_{t+1},X_t)$ can be written as 
       \begin{equation*}
       P(Y_{t+1},X_t) = P(Y_{t+1},X_t,Z_t \le R)+P(Y_{t+1},X_t,Z_t > R),
       \end{equation*}
       and by chain rule of probability distributions,
        \begin{align*}
       P(Y_{t+1},X_t) &= \frac{1}{2}[P(Y_{t+1}|X_t,Z_t\le R)P(Z_t\le R) +\\ &P(Y_{t+1}|X_t,Z_t >  R)P(Z_t > R)],  
       \end{align*}
       since $ X_t \perp Z_t$ and $P(X_t)=1/2$. 
       When $\lambda \le R$, $Y_{t+1}=X_t$ by the model setting so that  $\text{TE}_{X\to Y}(\lambda)$ is always unity. Thus in the following we focus on the case of $\lambda >R$. $P(Y_{t+1},X_t| Z_t\le \lambda)$ can be written as
        \begin{align*}
       P(Y_{t+1},X_t| Z_t\le \lambda) &= \frac{1}{2}[P(Y_{t+1}|X_t,Z_t\le R, Z_t\le \lambda)\times \\
       &P(Z_t \le R| Z_t\le \lambda)+\\
        &P(Y_{t+1}|X_t,Z_t  >  R,Z_t\le \lambda)\times \\
        &P(Z_t  > R| Z_t\le \lambda)].
       \end{align*}
Here the first and second terms in the right hand side of above equation, respectively, correspond to the case of the ``distance'' $Z_t$ being within the interaction ``radius'' $R$ ($Z_t \le R$) and that of $Z_t$ being larger than $R$ where there exists no interaction ($R < Z_t \le \lambda$). $P(Z_t \le R| Z_t\le \lambda)$ and $P(Z_t > R| Z_t\le \lambda)$ are equal to $\frac{R}{\lambda}$ and $1-\frac{R}{\lambda}$, respectively, since we have chosen a uniform distribution of $Z_t$.
      In the first term of the interaction regime,
       $P(Y_{t+1}|X_t,Z_t\le R,Z_t \le \lambda)$
       is divided into two cases,  one case where $Y_{t+1}, X_t$ are the same 
       and another
       case where $Y_{t+1}, X_t$ are different. 
       In the case that they are the same, $P(Y_{t+1}|X_t,Z_t \le R, Z_t\le \lambda) =P(Y_{t+1}|X_t,Z_t \le R)= 1$, and in the case where $Y_{t+1}, X_t$ are different, $P(Y_{t+1}|X_t,Z_t \le R, Z_t\le \lambda) =0$ by definition of the model. 
       In the second term of the no-interaction regime,
       $P(Y_{t+1}|X_t,Z_t>R,Z_t \le \lambda)=P(Y_{t+1})=\frac{1}{2}$ since $Y_{t+1} \perp  X_t$ for $Z_t>R$.
       
       The above equation implies, in the computation of $P(Y_{t+1},X_t| Z_t\le \lambda)$, that the contribution of independently-moving pairs of $Y_{t+1}$ and $X_t$ that do not ``interact'' to each other becomes dominated, as $\lambda$ gets larger than the interaction domain $R$.
       
       Finally in order to compute $\text{TE}_{X\to Y}(\lambda)$, we plug in \t{all the values} 
       back into  Eq.~\ref{eq: binary_TE_log} and obtain Eq.~\ref{eq: analytical_TE}. The function is differentiated with respect to
       $\lambda$ 
       to obtain Eq.~\ref{eq: analytical_dTE}.  
       Eq.~\ref{eq: analytical_TE} suggests that, as the cutoff distance $\lambda$ increases longer than the underlying interaction domain, $\text{TE}_{X\to Y}(\lambda)$ monotonically decreases (i.e., downward-convex) and converges to some nonzero finite value when the space size $L$ is finite (because $\lambda \le L$) otherwise zero. This further implies that conventional computation procedure of transfer entropy taking into account all pairs of agents, including non-interacting pairs, should ``dilute'' the contribution of interacting pairs in the system, which can yield some misleading interpretation for the relationship among the agents. 

      \bibliography{Reference}

\end{document}